\documentclass[fleqn,usenatbib]{mnras}

\usepackage{newtxtext,newtxmath}

\usepackage[T1]{fontenc}

\DeclareRobustCommand{\VAN}[3]{#2}
\let\VANthebibliography\thebibliography
\def\thebibliography{\DeclareRobustCommand{\VAN}[3]{##3}\VANthebibliography}

\usepackage{graphicx}	
\usepackage{amsmath}	

\newcommand\gaia {\textit{Gaia }}

\newcommand \grvs {$G_{\mathrm{RVS}}$}
\newcommand \sigrv {$\sigma_{\mathrm{RV}}$}

\title[Halo Stars Binary Fractions]{Close Binary Fractions in \emph{accreted} and \emph{in-situ} Halo Stars}

\author[Bashi et al.]{
Dolev Bashi,$^{1}$\thanks{E-mail: db975@cam.ac.uk}
Vasily Belokurov,$^{2}$
Simon Hodgkin$^{2}$\\
$^{1}$Astrophysics Group, Cavendish Laboratory, University of Cambridge, JJ Thomson Avenue, Cambridge CB3 0HE, UK\\
$^{2}$Institute of Astronomy, University of Cambridge, Madingley Road, Cambridge CB3 0HA, UK\\
}

\date{Accepted XXX. Received YYY; in original form ZZZ}

\pubyear{2024}

\begin{document}
\label{firstpage}
\pagerange{\pageref{firstpage}--\pageref{lastpage}}
\maketitle

\begin{abstract}
The study of binary stars in the Galactic halo provides crucial insights into the dynamical history and formation processes of the Milky Way. In this work, we aim to investigate the binary fraction in a sample of \emph{accreted} and \emph{in-situ} halo stars, focusing on short-period binaries. Utilising data from \gaia DR3, we analysed the radial velocity (RV) uncertainty \sigrv~distribution of a sample of main-sequence stars. We used a novel Bayesian framework to model the dependence in \sigrv~of single and binary systems allowing us to estimate binary fractions $F$ in a sample of bright (\grvs < 12) \gaia sources. We selected the samples of \emph{in-situ} and \emph{accreted} halo stars based on estimating the 6D phase space information and affiliating the stars to the different samples on an action-angle vs energy ($L_{\mathrm{z}}-E$) diagram. Our results indicate a higher, though not significant, binary fraction in \emph{accreted} stars compared to the \emph{in-situ} halo sample. We further explore binary fractions using cuts in $E$ and $L_z$, and find a higher binary fraction in both high-energy and prograde orbits that might be explained by differences in metallicity. By cross-matching our \gaia sample with APOGEE DR17 catalogue, we confirm the results of previous studies on higher binary fractions in metal-poor stars and find the fractions of \emph{accreted} and  \emph{in-situ} halo stars consistent with this trend. Our finding provides new insights into binary stars' formation processes and dynamical evolution in the primordial Milky Way Galaxy and its \emph{accreted} dwarf Galaxies.
\end{abstract}

\begin{keywords}
binaries: close -- Galaxy: halo -- Galaxy: kinematics and dynamics -- methods: statistical --
techniques: radial velocities
\end{keywords}



\section{Introduction}
The Galactic halo, a vast and sparsely populated region enveloping the Milky Way, presents a unique environment for the study of stellar populations \citep{Helmi2008,Helmi2020,DeasonBelokurov24}. It comprises predominantly old stars, remnants of both the primordial galactic formation era and the subsequent accretion events. This dichotomy in the halo's star formation history is epitomised in the two distinct groups of stars: `\emph{accreted}' halo stars, originating from smaller, consumed galaxies and the `\emph{in-situ}' halo stars, formed within the nascent Milky Way \citep{DeasonBelokurov24}. 

 The \emph{accreted} component of the halo in the solar neighbourhood, is exemplified by the \textit{Gaia}-Sausage-Enceladus \citep[GS/E;][]{Belokurov18, Helmi18, Haywood18}, a prominent substructure discerned through analyses of \gaia data \citep{GaiaDR1, Gaia18} indicating the remnant of the Milky Way's most consequential recent merger, an event estimated to have concluded around $8-10$ billion years ago. The GS/E appears to dominate the halo around the Sun, but a wide variety of smaller halo substructures have also been detected \citep{Helmi99, Myeong2018vel,Myeong2018act,Myeong19, Matsuno19, Koppelman2019,Naidu20,Ohare2020, Malhan2022, Dodd2023}.
 
 Conversely, the \emph{in-situ} halo population includes the `splashed' high-$\alpha$ disc \citep{Bonaca17, Gallart2019,diMatteo2019,Belokurov20}, and the {\it Aurora} stars \citep{BelokurovKravtsov22, Conroy2022,Chandra23, Zhang24}, both formed in the Milky Way proper before the arrival of the GS/E progenitor. \emph{Splash} stars, with their relative metal-rich composition ([Fe/H] $\gtrsim -1$) and highly eccentric orbits, are thought to have been displaced from the Galactic disc during the cataclysmic merger event that formed the GS/E structure. Having been formed before the merger, \emph{Splash} stars are on average as old as the stars in the GS/E \citep[see e.g.][]{Belokurov20}. \emph{Aurora} is characterized by its metal-poor nature ([Fe/H] $\leq$ -1) and a wide range of azimuthal velocities \cite[][]{BelokurovKravtsov22, Chandra23,Zhang24}. Identified through chemical tagging, \emph{Aurora} represents the oldest in-situ stellar population of the Milky Way, formed before the Galaxy had developed a coherently rotating disc \citep[see discussion in][]{Semenov2024, Dillamore2024}. Although first spotted in the vicinity of the Sun, \emph{Aurora} likely follows a steep radial density, with most of its mass near the Galactic centre \citep[see][]{BelokurovKravtsov23, Rix22}.

Alongside the \emph{accreted} stars, these components of the \emph{in-situ} halo, \emph{Splash} and \emph{Aurora},  underscore the Milky Way's dynamic history, from its earliest star formation episodes to the significant reshaping triggered by galactic mergers. This layered approach facilitated by the {\it Gaia} space observatory data, provides a nuanced understanding of how different stellar populations within the Galactic halo offer insight into the chronological sequence of events that shaped the Milky Way.

In the context of the Galactic halo, the study of binary stars acquires additional significance.
Binary stars are a fundamental component of our Galaxy's architecture. They serve as critical probes for understanding stellar formation and evolution \citep{CarneyLatham87, Tokovinin06, Latham02, Raghavan10, Moe19, Price-Whelan20, Mazzola20, Bashi23, NSS23, Chen24}. The frequency and characteristics of binary systems in this region provide insights into the conditions of star formation and the dynamical processes in the early Galaxy and the smaller, assimilated dwarf galaxies. For a recent review of results in the study of binary star research, particularly those facilitated by \textit{Gaia}, we refer the reader to \cite{ElBadry24}.

Previous studies have predominantly focused on binary systems within the disc and bulge of the Milky Way or have treated the halo as a homogeneous entity. One exception might be the recent work of \cite{Hwang22}, which focused on wide-binary stars fractions.

In recent years, the field of astrophysics has seen an influx of data from wide-field spectroscopic surveys such as LAMOST \citep{ Cui12}, GALAH \citep{GALAH15} and APOGEE \citep{MajewskiAPOGEE17}. These surveys have been instrumental in measuring the Radial-Velocities (RVs) and chemical abundances of hundreds of thousands of stars. A key aspect of some of these surveys is the collection of multi-epoch data for individual stars. This approach has enabled a statistical analysis of RV variability as a function of various stellar properties, including metallicity \citep{Tian18, Moe19, Price-Whelan20} and $\alpha$ elements \citep{Badenes18, Mazzola20, Niu21}.

For this study, we meticulously selected our sample from Data Release 3 \citep[DR3][]{Vallenari23}, focussing on specific criteria to ensure the relevance and accuracy of our analysis. While \gaia DR3 does not include each source's individual epoch RV measurements \citep{Katz23}, we can exploit the available RV variability information such as the standard deviation of the epoch RV measurements of the individual source \sigrv, or the $\Delta \mathrm{RV_{max}}$, the maximum detected shift in the radial velocities to assess binary frequency in a statistical framework \citep{Maoz12, Badenes18, Moe19, Mazzola20, Andrew22}. 

 
 While it is generally hypothesised that \emph{accreted} and \emph{in-situ} halo stars might exhibit different binary fractions due to their distinct formation histories, empirical evidence supporting this notion is sparse. This research aims to fill this gap by employing a novel methodological approach to estimate the close binary fractions among these stars, thereby shedding light on the diverse evolutionary pathways that have shaped the Galactic halo.

The paper is structured as follows. In Section~\ref{sec:MsSample}, we present our sample selection of bright \gaia Main-Sequence (MS) sources. Section~\ref{sec:stellarvari} discusses the dependence between \sigrv~and the stellar magnitude seen in the \gaia sample. In Section~\ref{sec:haloSample}, we demonstrate our selection procedure of \emph{accreted} and \emph{in-situ} halo stars. 
Section~\ref{sec:Methodology} presents our methodological approach using Bayesian modelling of the binary fractions. Sections~\ref{sec:halofrac} and \ref{sec:Metallicityfrac} give our main results on the binary fractions in the \emph{accreted} and \emph{in-situ}, as well as their dependence on metallicity. Section~\ref{sec:Conclusions} summarises our results and discusses their possible implications.

\section{selecting a sample of bright main-sequence sources}
\label{sec:MsSample}

We targeted stars within $3$ kpc of the solar vicinity with measurable radial velocities \citep{Katz23}. The query used to obtain data from the \gaia DR3 was designed to filter and isolate a specific subset of MS and sub-Giant objects, focussing on those relevant to our study of binary fractions in the Galactic halo. Consequently, we limited our sample to stars with magnitudes estimated from the integration of the Radial Velocity Spectrometer (RVS) \grvs~brighter than $12$ mag using \texttt{rv\_method\_used=1}. \grvs~is estimated using the flux in the RVS spectra \citep{Sartoretti23}. This magnitude constraint was applied to ensure the reliability and quality of the RV data, as the epoch radial velocities of stars beyond this threshold were not considered reliable enough to derive the combined radial velocities as the median of the epoch radial velocities \citep{Katz23}.

We further selected stars with well-constrained parallax measurements (\texttt{parallax\_over\_error} > 5 and \texttt{parallax} > 0) , ensuring high-quality distance estimates. 
To minimise contamination from disc stars, we selected stars with a galactic latitude ($b$) satisfying the condition $|b|>10 ^{\circ}$. This criterion was essential to focus our study on halo stars and reduce the inclusion of stars from the galactic disc with different dynamical properties.

We used \gaia extinction values to accurately plot the Colour-Magnitude Diagram (CMD) of our sample. The inclusion of stars with non-null extincation values ensure that we have reliable photometric extinction-corrected data, crucial for accurate luminosity and temperature determination. Specifically, we corrected the G-band magnitudes (Abs. G) and BP-RP colours using the \texttt{ag\_gspphot} and \texttt{ebpminrp\_gspphot} values, respectively, to accurately represent the absolute magnitudes on the Y-axis and the BP-RP colours on the X-axis of the CMD plot.

Within the CMD, we employed a polygonal selection method to isolate mainly dwarf stars. Our decision to focus on MS stars stemmed from our objective of studying stellar variability caused by motion in a binary system. Using this selection, we avoided the added complexity in cases of evolved stars with their variable and evolutionary transient stages as well as the impact of stellar evolution on binary fraction.

To further constrain our sample to select stars that were observed with good coverage in epoch phase-space, we selected sources with \texttt{rv\_visibility\_periods\_used} larger than 8 where a visibility period is a group of transits separated from other groups by a gap of at least $4$ days \citep{Vallenari23, Katz23}.

\begin{figure*}	
\includegraphics[width=18cm]{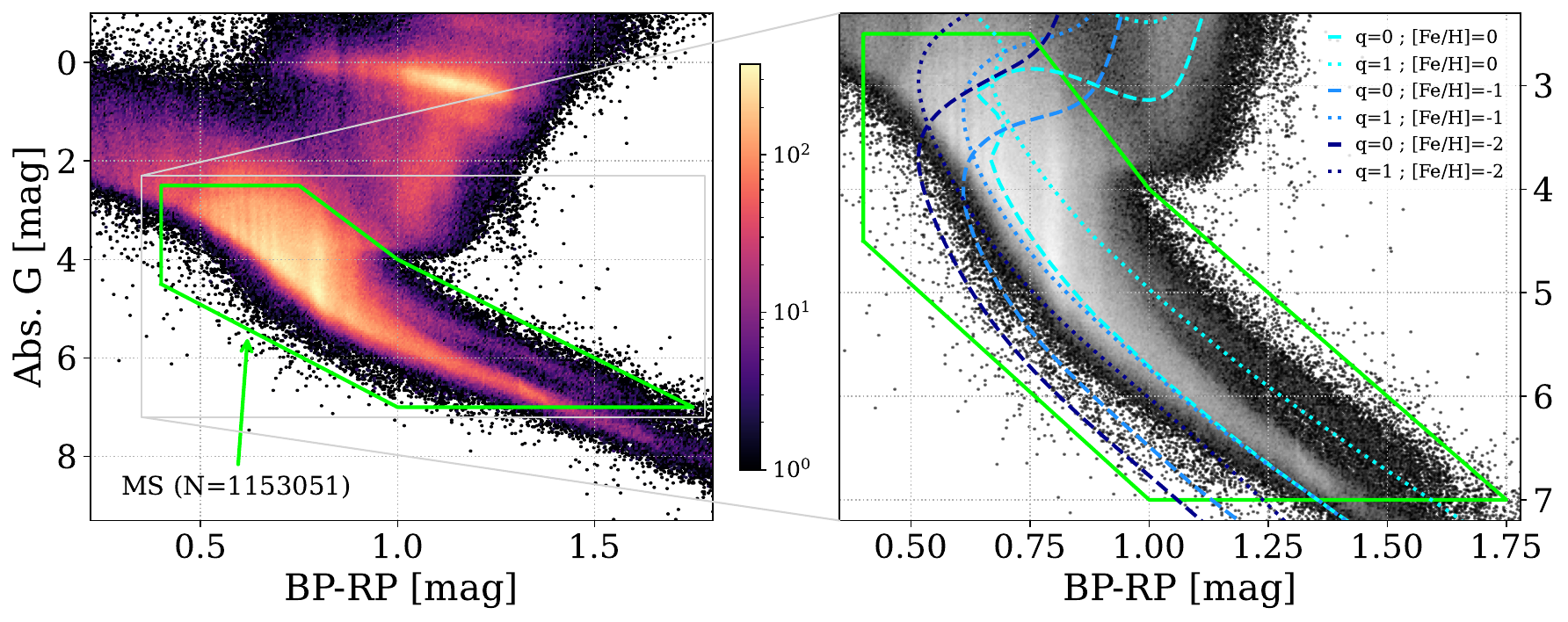}
    \caption{Extinction corrected Colour-Magnitude Diagram (CMD) of available \gaia sources with combined radial velocities. Left panel displays points density with colour scale represents the number of sources falling into each bin. The green polygon-shaped dashed line (see text) delineates our border of the selected MS sample. Right panel shows a zoomed-in view with MIST Isochrone for single (q=0) and twin binary (q=1) systems in different range of metallicities and ages.}
    \label{fig:cmd}
\end{figure*}

In the left panel of Fig.~\ref{fig:cmd}, we show a CMD of our sample of $N=1,153,051$ MS and Sub-Giants sources (out of a total stellar sample of $N=2,174,836$ sources) delineated by green polygonal borders.\footnote{The used polygon bounds are the following (\texttt{BP\_RP}, \texttt{Abs. G}): (0.4, 4.5), (0.4, 2.5), (0.75, 2.5), (1, 4), (1.75, 7), (1, 7), (0.4, 4.5).} The colour scale represents the number of sources falling into each bin. The right panel of Fig.~\ref{fig:cmd} displays a zoomed selection of the polygonal area, where we have superimposed the MESA \citep{MESA} Isochrones and Stellar Tracks models \citep[MIST ;][]{MIST0, MIST1} Isochrones interpolation for single stars (i.e.  binary mass ratio, $q=0$) and twin binaries ($q=1$). For single stars, the tracks span a range of metallicities, accurately tracing the progression of metal-poor to metal-rich stars within the MS. For the [Fe/H]=-1 and [Fe/H]=-2 a $12$ Gyrs Isochrone was selected while for the [Fe/H]=0 we selected a $4.5$ Gyrs Isochrone to better characterise the current population of stars in these metallicity bins. Additionally, for the case of twin binaries, the corresponding track is plotted $0.75$ magnitude above the single star line \citep{HurleyTout97} representing the photometric appearance of an equally bright companion in the CMD, effectively mimicking the luminosity of a normal metal-rich dwarf. 

As can be seen by Fig.~\ref{fig:cmd}, the polygon borders are meticulously chosen to include mainly dwarf stars and capture the sequence of metal-poor stars, typically found at the lower extremity of the MS. Notably, the binary sequence of metal-rich stars (at \texttt{BP\_RP} > 1) is clearly discernible within our selection and corresponds nicely to the twin stars [Fe/H] = 0 Isochrone. While most stars in this region are not halo stars, their inclusion was deemed essential for completeness. As a final remark, we note that in the case of a metal-poor star with an equally bright companion (twin) the photometrically position on the CMD, appear as a normal metal-rich dwarf.

\section{Stellar variability: radial velocity uncertainty}
\label{sec:stellarvari}
To get the standard deviation of the epoch RV measurements of the individual bright source \sigrv, we used the following expression:
\begin{equation}
\sigma_{\mathrm{RV}} =\sqrt{\frac{2N_{\mathrm{RV}}}{\pi}\left(\zeta_{\mathrm{RV}}^2 - 0.11^2 \right)}
	\label{eq:sigmrv}
\end{equation}
where $\zeta_{\mathrm{RV}}$ is the uncertainty on the median of the
epoch radial velocities (\texttt{radial\_velocity\_error}) to which a constant shift of $0.11$~km/s was added to take into account a contribution from the calibration floor \citep{Andrew22, Katz23}, and $N_{\mathrm{RV}}$ is the number of transits used to derive the median RV (\texttt{rv\_nb\_transits}). 

\begin{figure}
	\includegraphics[width=\columnwidth]{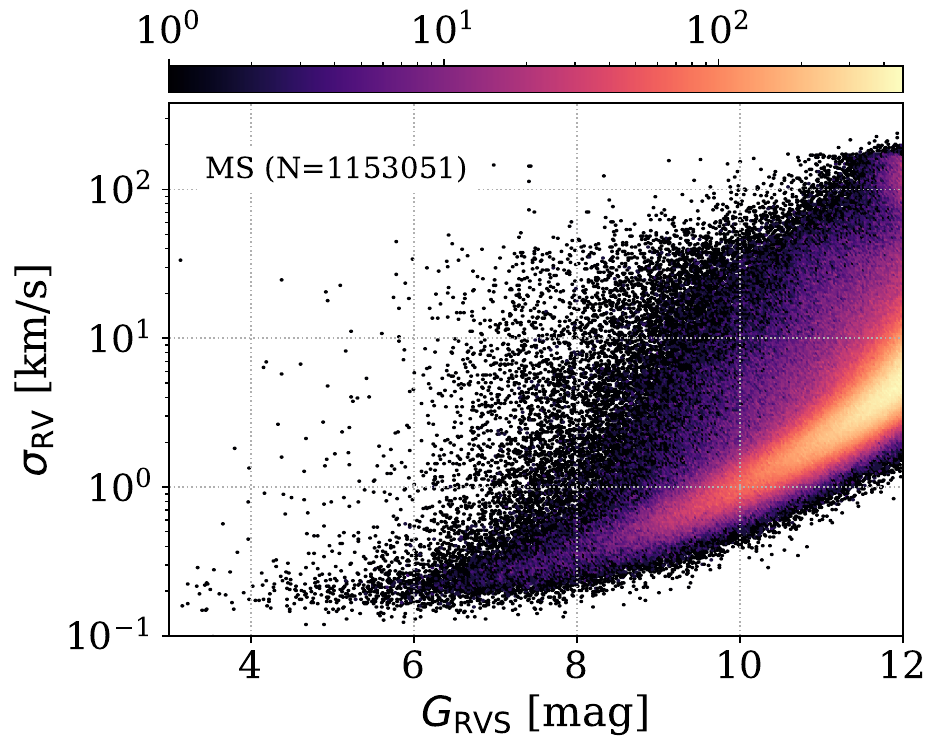}
    \caption{RV uncertainty \sigrv~vs \grvs~of our MS sample (sources inside green polygon of Fig.~\ref{fig:cmd}). The colour scale represents the number of sources falling into each bin. A clear monotonic trend between \sigrv~and \grvs~is evident, with points well above this trend being suspected binary systems.}
    \label{fig:grvs_sig}
\end{figure}

Fig.~\ref{fig:grvs_sig} shows the distribution of $N=1,153,051$ MS sources in \sigrv~as function of \grvs. 
As might be expected, the typical standard deviation of the epoch RV measurements increases towards fainter sources. 
Above this observable curve, we find a wider spread of sources of suspected binary systems with high \sigrv~values. 

\begin{figure}
	\includegraphics[width=\columnwidth]{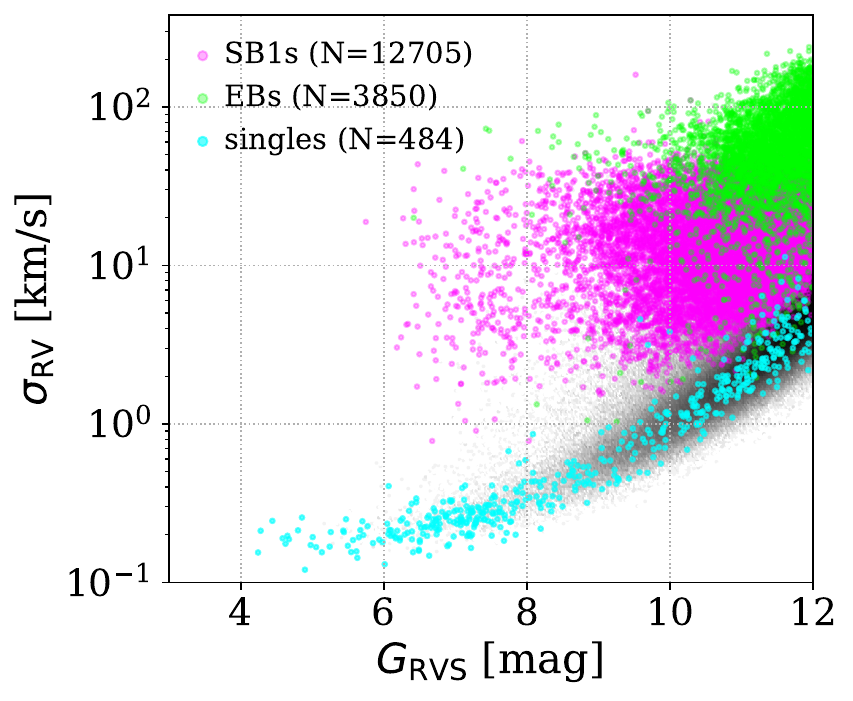}
    \caption{RV uncertainty \sigrv~vs \grvs~of single stars (cyan), single-line spectroscopic binaries listed in the \gaia DR3 \textit{NSS} catalogue (magenta) and eclipsing binaries listed in the first \gaia catalogue of eclipsing-binary candidates (green) after cross match with our MS sample. Gray background colour scale represents the number of sources falling into each bin of the general sample similar to Fig.~\ref{fig:grvs_sig}.}
    \label{fig:grvs_sig_single_binary}
\end{figure}

To demonstrate this expected trend, we show in Fig.~\ref{fig:grvs_sig_single_binary} the \sigrv-\grvs~scatter of a sample of three types of systems: (i) fraction of planet-host stars listed in the NASA Exoplanet Archive\footnote{https://exoplanetarchive.ipac.caltech.edu/index.html} with a flag suggesting detection by radial-velocity and can be approximated as a well-constrained sample of single stars; (ii) a sample of single-lined spectroscopic binaries (SB1) listed in the \gaia DR3 \textit{NSS} catalogue \citep{NSS23} with a \texttt{clean-score} above 0.578 based on \cite{Bashi22} metric and (iii) eclipsing binaries (EB) listed in the first \gaia catalogue of EB candidates \citep{GaiaEB}. As expected, the single stars sample sits nicely on the exponential curve and have the lowest \sigrv~for a given \grvs, while the SB1s are found above this curve, and the EBs with typically shorter periods and larger semi-amplitudes have the highest \sigrv~values.

In Section~\ref{sec:Methodology}, we will further elaborate on this dependence in developing our model to estimate binary fractions and the typical (mean) \sigrv~value as a function of \grvs.  


\section{\emph{accreted} and \emph{in situ} samples}
\label{sec:haloSample}
In addition to the criteria mentioned above, we implemented further cuts to refine our selection of halo stars.
Specifically, we applied a velocity threshold, selecting stars with a total velocity ($V_{\mathrm{tot}} = \sqrt{U^2 + V^2 + W^2}$) greater than $220$ km/s. This threshold is based on the understanding that the halo shows close to net zero rotation, and therefore in the Solar neighbourhood, the velocity of its stars is on average offset from that of the disc's stars by the local value of the circular velocity \citep{NissenSchuster10, Helmi18, Belokurov18,  BashiZucker22, Lane22}.

To accurately estimate the dynamical properties of our selected stars, we used the \texttt{Galpy} package \citep{Bovy15}. Using \texttt{Galpy}, we calculated the orbital energy ($E$) and the z-component of angular momentum ($L_z$) for each star in our sample. 
These calculations were based on the assumption of the Milky Way's potential as described in the \texttt{MWPotential2014} model, which is a well-established model that approximates the gravitational potential of the Milky Way, considering the contributions from the bulge, disc, and halo. We plot all energies offset by the potential of \texttt{MWPotential2014} at infinity so that $E<0$ are bound orbits \citep[e.g.][]{Lane22}. 

In the $E-L_z$ plane, the
samples are confined by a bounding parabola defined by the escape velocity as a function of the change in energy at a fixed Galactocentric radius. 

\begin{figure*}
	\includegraphics[width=13.5cm]{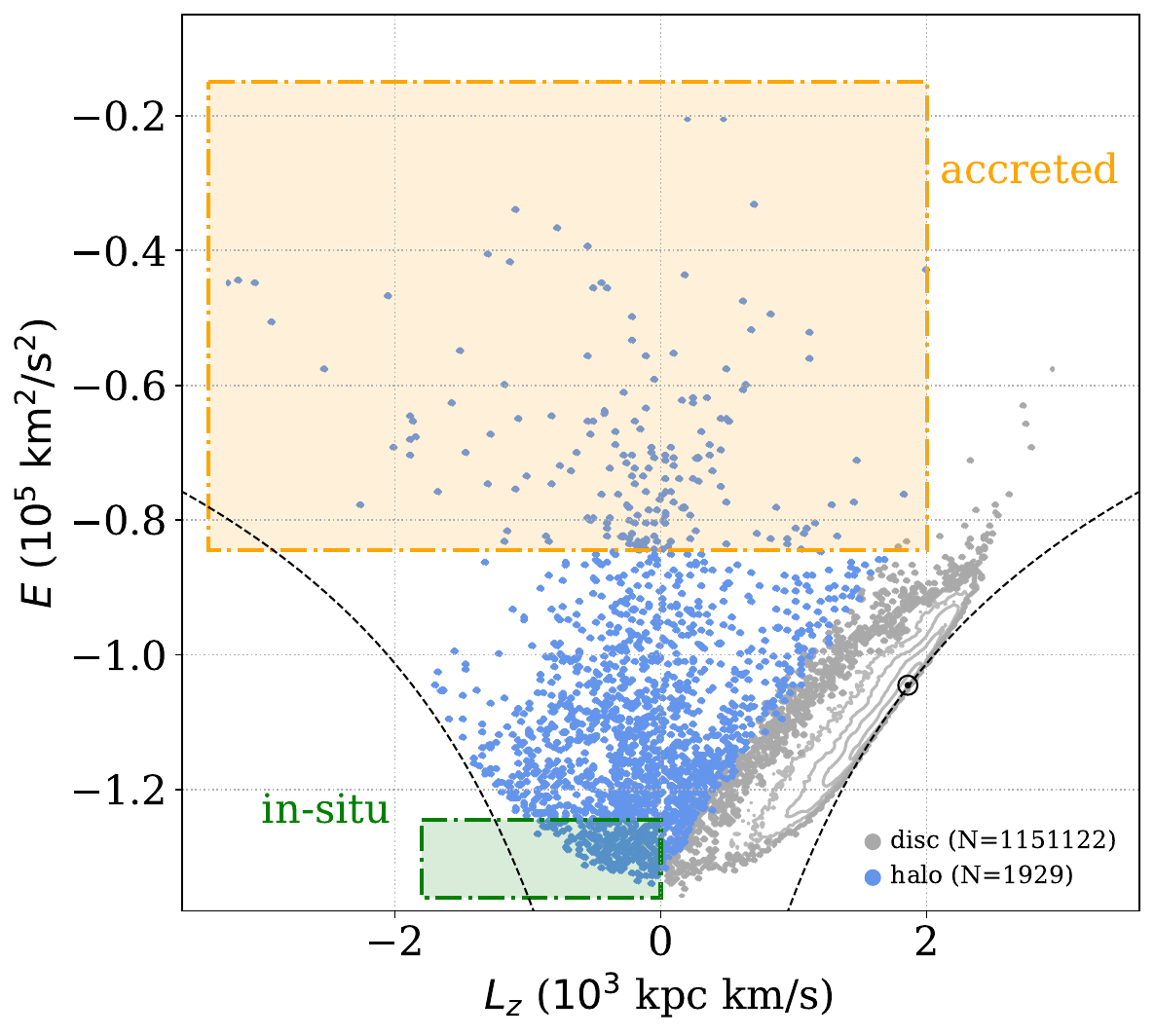}
    \caption{$E-L_z$ values of our MS \gaia sources, with contour background grey points mark the main location of disc stars and blue points mark stars with $V_{\mathrm{tot}} > 220$ km/s (see text for full selection cuts). Blue points inside the dashed green box mark our selected sample of '\emph{in-situ}' sources, while blue points inside the dashed orange box mark the sample of \emph{accreted} stars. The black fine-dashed curve corresponds to the maximal angular momentum at fixed energy, i.e. orbits with circular velocity. $\odot$ marks the location of the Sun in the chosen potential.}
    \label{fig:Lz_E}
\end{figure*}

In Fig.~\ref{fig:Lz_E} we show $N=1,928$ stars on the $E-L_z$ plane that were left after applying the above cuts and a background grey contours marking the position of the overall MS sample which mainly consisted of disc stars. We mark the Sun's position at $E_{\odot}=-1.04 \times 10^5~\mathrm{km^2/s^2}$ and use it to gauge the boundary energy we select to separate between the \emph{accreted} and \emph{in-situ} samples. Follow \cite{BelokurovKravtsov23}, and to reduce contamination, we selected conservative cuts in the $E$ and affiliated \emph{accreted} stars as sources with $E > E_{\odot} + 0.2$ and $L_z < 2$ and \emph{in-situ} stars as $E < E_{\odot} - 0.2$ and $L_z < 0$. 

By incorporating these velocity and dynamic cuts, we were able to identify stars that are characteristic of the Milky Way’s halo more precisely and were left with $N=181$ \emph{accreted} and $N=291$ \emph{in-situ} halo stars. This refined selection ensures that our analysis of binary fractions is focused on the intended stellar population, thereby enhancing the reliability of our binary fractions estimates.

\section{Methodology: Modelling Binary Fractions using the RV uncertainty distribution} 
\label{sec:Methodology}
Single stars typically exhibit minimal RV variation, primarily due to factors such as stellar activity and instrumental noise, resulting in low \sigrv~values that follow a trend similar to the cyan points of Fig.~\ref{fig:grvs_sig_single_binary} with fainter stars having a larger variation. 
On the other hand, in binary systems, the orbital motion of each star causes periodic changes in their radial velocities. For a given binary semi-amplitude, $K$, in the most simplified scenario of a circular orbit, assuming that the measurements are uniformly distributed across the orbital phase, the added RV uncertainty from the binary nature of the system would be $\sigma_{\mathrm{RV}}^b=K/\sqrt{2}$. Thus, the standard deviation of the value of the epoch RV measurements is a direct indicator of the binary nature of the system \citep{Maoz12,Badenes18,Moe19,Mazzola20, Andrew22}, where a larger \sigrv~typically suggests a closer or more massive companion as $K \propto P^{-1/3} q  (1+q)^{-2/3}$ where $P$ is the binary orbital period and $q$ is the binary mass ratio.

By analysing the \sigrv-\grvs~distribution for a given sample of \gaia stars, we can statistically estimate the binary fraction. Given a typical \sigrv~of $\sim 5$ km/s at \grvs=12, our method is sensitive to close binary systems with orbital period $P \lesssim 1000$ days for a solar mass twin binary system. Nonetheless, it important to note that detecting twin sources with mass ratios close to one ($q \simeq 1$) using \gaia RVS is more challenging because the Doppler shifts of the two similar stars partially cancel each other out in the combined spectrum. This cancellation reduces the observable RV variability, and consequently the \sigrv~value. 

A dataset consisting predominantly of single stars would exhibit a narrow distribution centred around and following the trend of dense sources shown in Fig.~\ref{fig:grvs_sig}, thereby reflecting minimal RV variation. In contrast, a dataset with a high binary fraction would show a broader distribution, extending towards higher \sigrv~values for a given \grvs, indicative of the significant RV changes caused by binary companions. A mixed population of single and binary stars would likely display a bimodal distribution, with two distinct peaks corresponding to the two populations.

In most cases, our method treats higher-order multiple-star systems, such as triples, as binary stars. In this scenario, we might be sensitive to the close binary substructure where the ratio of the outer period $P_{\mathrm{out}}$ to the inner period $P_{\mathrm{in}}$, adjusted for eccentricity $e_{\mathrm{out}}$, is expected to have $P_{\mathrm{out}}\left( 1-e_{\mathrm{out}} \right)^3/P_{\mathrm{in}} > 5$ \citep{Tokovinin04} while we may miss the RV variation of the outer companion.

To model the population of sources in log \sigrv, we define a density function, which is a sum of two Gaussian distributions to model the populations of single and binary stars. To properly model the fractions of single and binary stars, we multiply each Gaussian by the expected fractions, where $F$ marks the binary fraction. 
The final density function will then be

\begin{equation}
    \begin{aligned}
        f(x | G_{\mathrm{RVS}}; \theta) &= (1 - F) \cdot \mathcal{N}_s\left(x | \mu_s(G_{\mathrm{RVS}}), \sigma_s^2\right)\\
        &+ F \cdot \mathcal{N}_b\left(x | \mu_b(G_{\mathrm{RVS}}), \sigma_b^2\right) .
    \end{aligned}
	\label{eq:pdf}
\end{equation}
where $\theta = {\left(F, a, b, G_{\mathrm{min}}, d, \sigma_s, \sigma_b\right)}$, defines our parameter space and $\mathcal{N}_s\left(x | \mu_s(G_{\mathrm{RVS}}), \sigma_s^2\right)$, $\mathcal{N}_b\left(x | \mu_b(G_{\mathrm{RVS}}), \sigma_b^2\right)$ are the Gaussian density functions of the single and binary populations with means $\mu_s(G_{\mathrm{RVS}})$, $\mu_b(G_{\mathrm{RVS}})$ and standard-deviations 
$\sigma_s$, $\sigma_b$ respectively. 

To model the dependence between \sigrv~and \grvs, we used an exponent function and defined $\mu_s(G_{\mathrm{RVS}})$ as
\begin{equation}
\mu_s(G_{\mathrm{RVS}}; a, ,b, G_{\mathrm{min}}) = a + e^{b\cdot(G_{\mathrm{RVS}} - G_{\mathrm{min}})} 
\label{eq:mu1}
\end{equation}
where $a,~ b,~ G$ are free parameters which control the shape of the expotential function. Similarly, we defined the binary mean, $\mu_b(G_{\mathrm{RVS}})$  in quadrature as
\begin{equation}
\mu_b(G_{\mathrm{RVS}}; a, ,b, G_{\mathrm{min}}, d) = \log_{10} {\left( \sqrt{\left(10^{\mu_s(G_{\mathrm{RVS}}; a, ,b, G_{\mathrm{min}})}\right)^2 + d^2}\right)}\\
\label{eq:mu2}
\end{equation}
where $d$ is a free parameter that accounts for the extra variability in \sigrv~as a result of the system's binary nature.

In our analysis, we sought to account for potential variations in \sigrv~attributed to line-broadening where early type stars at a fixed \grvs~having larger errors \citep{Andrew22}. To explore this dependence, we extend our model by introducing an additional factor to account for the relation between \sigrv~and \texttt{BP\_RP}. 
Despite this, the outcomes from our `simplified' model were found to align with those from a model that explicitly incorporates the spectral type. This agreement suggests that the parameters $\sigma_s$ and $\sigma_b$ within our model effectively encapsulate the influence of spectral type, thereby moderating the dispersion of data points around the fitted single- and binary-star models. 

\begin{table}
	\centering
	\caption{Prior distributions of model parameters of binary fraction ($F)$, and single and binary Gaussian's means ($a,~b,~G_{\mathrm{min}},~d$) and stds ($\sigma_s,~\sigma_b$). }
	\label{tab:priors}
	\begin{tabular}{lc} 
		\hline
		Parameter & Prior \\
		\hline
		$F$ & $\log \mathcal{U}(0.01,1)$\\
		$a$ & $\mathcal{U}(-5,5)$\\
		$b$ & $\log \mathcal{U}(10^{-3},1)$\\
  		$G_{\mathrm{min}}$ & $\mathcal{U}(0,12)$\\
    	$d$ & $\log \mathcal{U}(10^{-4},100)$\\
      	$\sigma_s$ & $\log \mathcal{U}(10^{-3},10)$\\
            $\sigma_b$ & $\log \mathcal{U}(10^{-2},10)$\\

		\hline
	\end{tabular}
\end{table}

Finally, since stars whose combined RV formal uncertainty was greater than $40$ km/s were flagged as invalid and removed from the final \gaia DR3 catalogue \citep{Katz23}, we used a truncated normal distribution to cope with this abrupt cut in the observed distributions.

The final likelihood expression would then be

\begin{equation}
\mathcal{L} = \prod_{i=1}^{N} f(x^i | G_{\mathrm{RVS}}^i; \theta) .
\label{eq:likelihood}
\end{equation}

\begin{figure}
 	\includegraphics[width=\columnwidth]{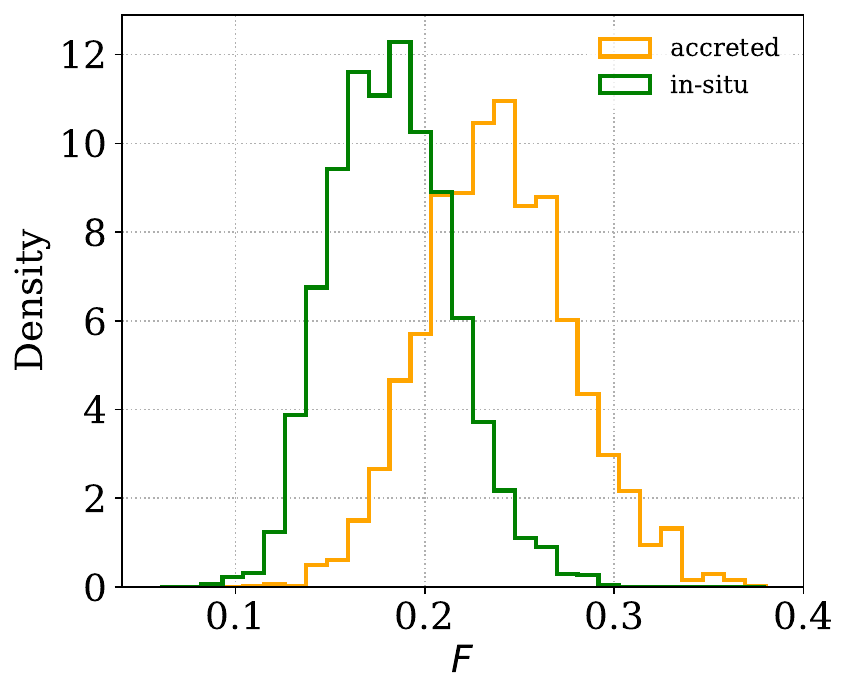}
    \caption{Posterior histograms of binary fractions in \emph{accreted} (orange) and \emph{in-situ} (green) samples based on the selection presented in section~\ref{sec:haloSample}.}
    \label{fig:Frac_accreted_in_situ}
\end{figure}

To test our algorithm, we applied it on different sets of simulated datasets and used an MCMC routine with $40$ walkers and $10^4$ steps, using uninformative priors on the seven free parameters of our model as listed in Table \ref{tab:priors}. We used the Python \texttt{emcee} package \citep{Foreman-Mackey13} to find the parameter values (and their uncertainties) that maximise the sample likelihood.

In the first test set, we attempted to find the best-fit model parameters ${\left(F, a, b, G_{\mathrm{min}}, d, \sigma_s, \sigma_b\right)}$ from a simulated sample based on the defined density function of Equation~\ref{eq:pdf}. 

In the second test set, we simulated a population of single vs. short-period binary systems using a random selection of binary separations, mass ratios, eccentricity distribution inclination and phase \citep[e.g.][]{Mazzola20} and attempted to get the best-fit parameters of ${\left(F, a, b, G_{\mathrm{min}}\right)}$. 

In the third test set, we used a mixed sample of known single stars and \gaia SB1s \citep{Bashi22, NSS23} based on the subsamples used in Fig.~\ref{fig:grvs_sig_single_binary} and attempted to get the best-fit parameters of ${\left(F\right)}$. 

We found good agreement between the posterior distributions and the actual simulated parameters in all test cases, thereby supporting our Bayesian model and encouraging us to proceed and estimate binary fractions on real datasets. 

\section{Binary fractions in halo stars}
\label{sec:halofrac}

\subsection{The \emph{accreted} and \emph{in-situ} subsamples}
\label{sec:acc_situfrac}

Given our selection of $N=181$ and $N=291$ \emph{accreted} and \emph{in-situ} sources, respectively, we list in Table~\ref{tab:posteriors} the best fit posterior values of our Bayesian model (median and $\%16, \%84$ quantiles uncertainties).
We present in Fig.~\ref{fig:Frac_accreted_in_situ} the posterior distributions of the \emph{accreted} (orange) and \emph{in-situ} (green) binary fractions. 
We find a $\sim2\sigma$ difference in favour of a higher binary fraction in \emph{accreted} stars, $F_{\mathrm{accreted}}=0.238\pm0.038$, compared to \emph{in-situ} stars, $F_{\mathrm{in-situ}}=0.182\pm0.032$. 

\begin{figure*}
    \centering
    \includegraphics[width=0.48\textwidth]{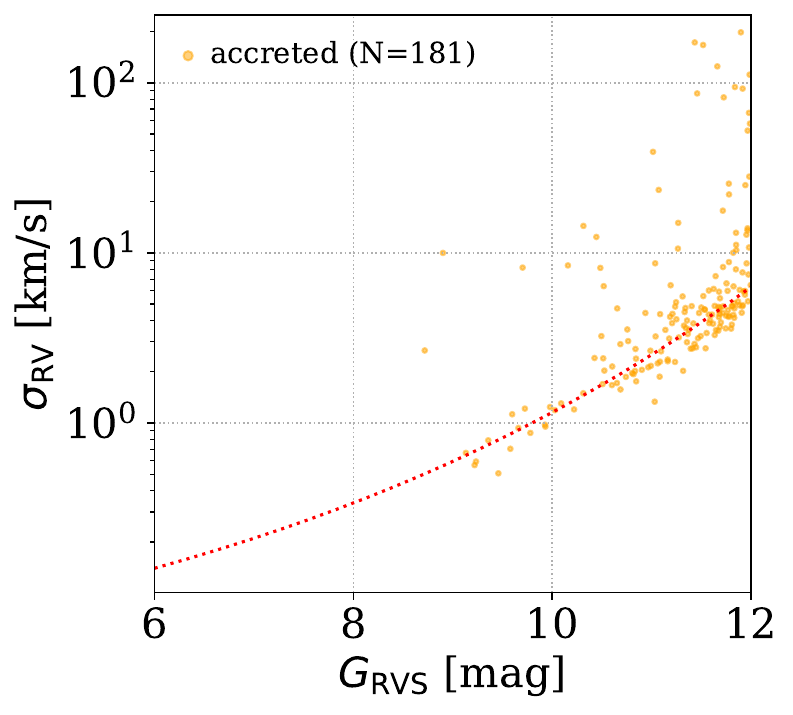}
    \includegraphics[width=0.48\textwidth]{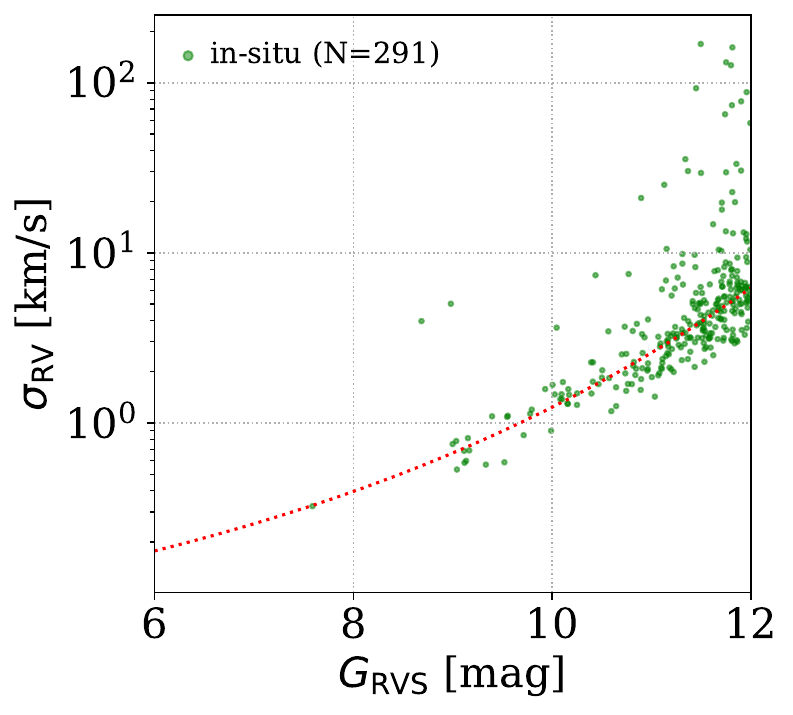}
    \caption{\sigrv~vs \grvs~of the \emph{accreted} (left panel) and \emph{in-situ} (right panel) samples. The dotted red lines mark the best-fit Bayesian model of our posterior distribution based on Equation~\ref{eq:mu1}.}
    \label{fig:sigrv_grvs_accreted_in_situ}
\end{figure*}

In Fig.~\ref{fig:sigrv_grvs_accreted_in_situ}, we show a scatter plot of \sigrv~vs. \grvs~ for the \emph{accreted} and \emph{in-situ} samples. The red dotted lines mark our best-fit posterior distributions of the mean values using Equation~\ref{eq:mu1} which describes the relation between \sigrv~and \grvs~in the case of a single source solution. In general, the best-fit posteriors ($a$, $b$, $G_{\mathrm{min}}$), which determine the shape of the curves of the \emph{accreted} and \emph{in-situ} samples, are consistent within $1\sigma$ as can be seen in Table.~\ref{tab:posteriors}. 

\begin{figure}
 	\includegraphics[width=\columnwidth]{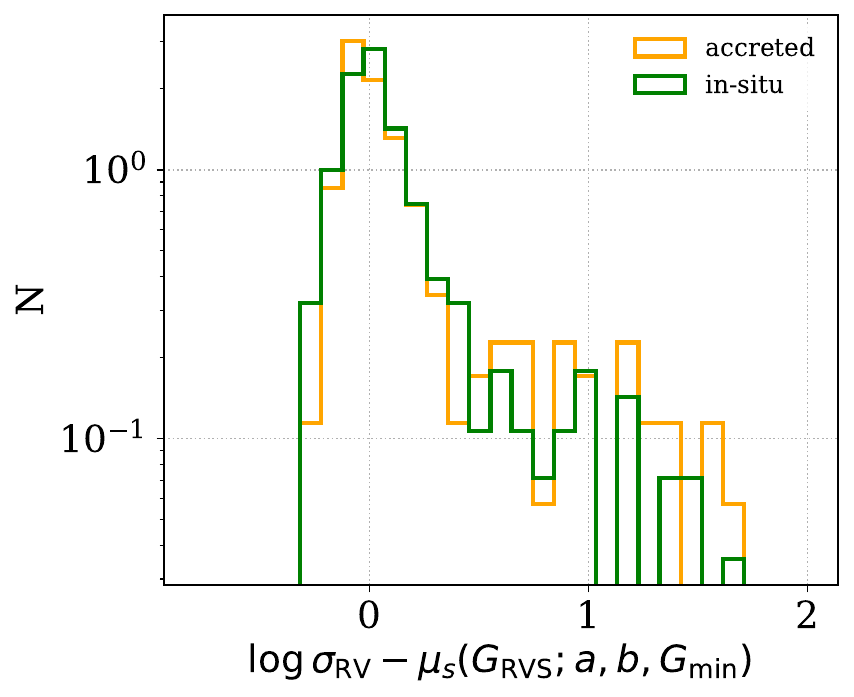}
    \caption{`Flattened' normalised distributions of \emph{accreted} (orange) and \emph(in-situ) (green) sources after subtracting the observed \sigrv~values with the best fit model $\mu_s(G_{\mathrm{RVS}}; a, b, G_{\mathrm{min}})$.}
    \label{fig:flat_acc_in-situ}
\end{figure}

To explore the distributions of \sigrv~after removing the effect of stellar \grvs, we show in Fig.~\ref{fig:flat_acc_in-situ} the `flattened' normalised distributions after subtracting the observed \sigrv~values with the best fit model $\mu_s(G_{\mathrm{RVS}}; a, b, G_{\mathrm{min}})$. Comparing these distributions might also shed some light on the nature of these binary systems. As expected, both histograms have clear peaks around zero, marking the distributions of single sources with standard deviation $\sigma_s$. The extended tail of binaries is also evident where we find the \emph{accreted} sample having more sources at high $\sigma_{\mathrm{RV}} - 
\mu_s(G_{\mathrm{RVS}}; a, b, G_{\mathrm{min}})$ values compared to the \emph{in-situ} sub-sample. This observed trend is also supported by our posterior model results where the \emph{accreted} best-fit model on $d$ and $\sigma_b$, which are responsible to mark the position and spread of the binary sequence, is larger than in the \emph{in-situ} case.   


A significant limitation encountered in using the \gaia sample relates to the obscured insights into the metallicity content of the two halo samples. Metallicity is known to anti-correlate with binary fraction \citep{Badenes18, Moe19, Price-Whelan20, Mazzola20}. Consequently, any disparity in the metallicity content between the \emph{accreted} and \emph{in-situ} samples can lead to variations in binary fractions that are not inherently related to their dynamical properties. To further explore this, we examine in Section \ref{sec:Metallicityfrac} a subset of our samples with reported metallicity estimates using the APOGEE DR17 catalogue \citep{APOGEE17}. 

\begin{table}
	\centering
	\caption{Posterior values of \emph{accreted} and \emph{in-situ} samples. }
	\label{tab:posteriors}
	\begin{tabular}{lcc} 
		\hline
		Parameter & \emph{accreted} & \emph{in-situ}\\
		\hline$N$ & $181$ & $291$ \\
$F$ & $0.2384 \pm 0.038$ & $0.182 \pm 0.032$ \\
$a$ & $-1.84_{-0.55}^{+0.44}$ & $-1.59_{-0.39}^{+0.45}$ \\
$b$ & $0.163_{-0.027}^{+0.059}$ & $0.174_{-0.034}^{+0.042}$ \\
$G_{\mathrm{min}}$ & $6.0_{-2.3}^{+2.5}$ & $7.0_{-2.4}^{+1.8}$ \\
$d$ & $20.0_{-5.0}^{+8.2}$ & $14.6_{-3.2}^{+4.0}$ \\
$\sigma_s$ & $0.1150_{-0.0091}^{+0.0094}$ & $0.1252_{-0.0082}^{+0.0092}$ \\
$\sigma_b$ & $0.601_{-0.075}^{+0.097}$ & $0.519_{-0.055}^{+0.066}$ \\
		\hline
	\end{tabular}
\end{table}

\subsection{Cuts in $E$ and $L_z$}
\label{sec:E_Lz_Frac}

We next attempted to estimate binary fractions using different cuts in $E$ and $L_z$ that might be associated with different Galactic components \citep{HelmiWhite99, Belokurov18, Myeong19, Matsuno19, Naidu20, BelokurovKravtsov22, BelokurovKravtsov23}. 
While \emph{accreted} stars and especially the GS/E, are more common at higher $E$ values in the solar neighbourhood \citep{Belokurov18, Helmi18, Haywood18,Naidu20}, other substructures also have their expected places on the $E-L_z$ diagram. For example, at high $L_z$, we are evident to the extended high-$\alpha$ sequence of disc stars as well as \emph{Splash} stars \citep{Bonaca17, Gallart2019,diMatteo2019,Belokurov20}. This sample might be contaminated by a higher fraction of metal-rich stars. Similarly, at low energy values of retrograde orbits, we should expect to have a higher population of \emph{Aurora}, \emph{in-situ} stars \citep{BelokurovKravtsov22, Conroy2022,Chandra23, Zhang24}. 

\begin{figure}
	\includegraphics[width=0.48\textwidth]{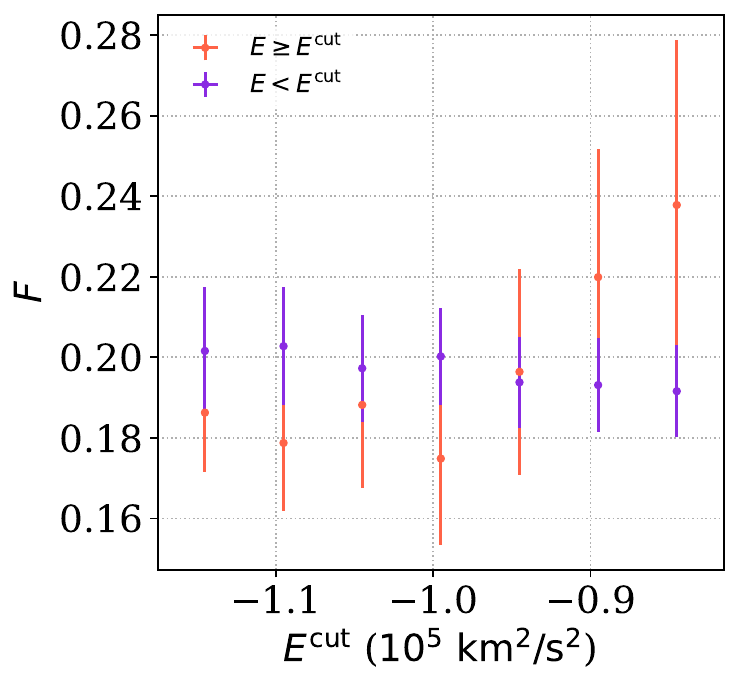}
 \includegraphics[width=0.48\textwidth]{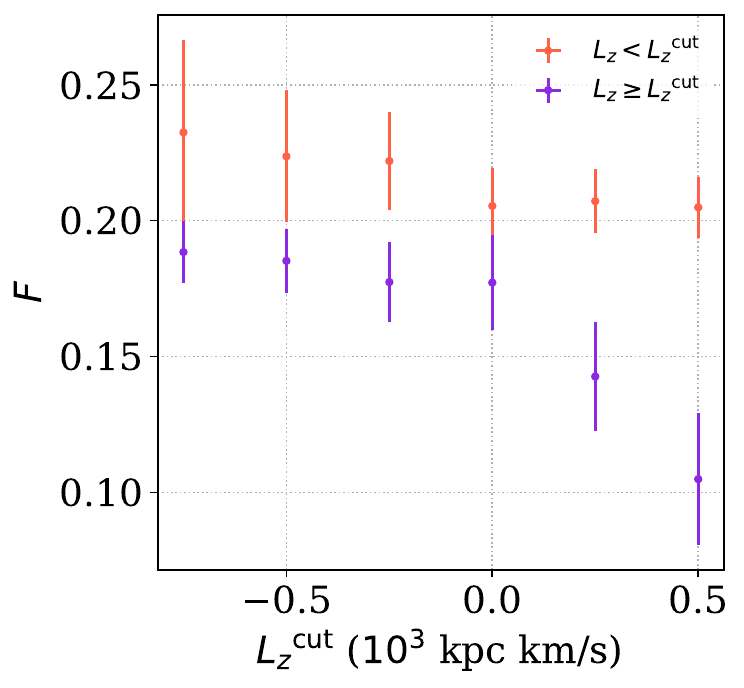}
    \caption{Binary fractions in halo stars and their corresponding uncertainties using different cuts in energy ($E^{\mathrm{cut}}$, upper panel) and angular momentum ($L_z^{\mathrm{cut}}$, bottom panel). Red points display sources with $E$ ($L_z$) higher (lower) than the threshold, while violet points display the complement subsets.}
    \label{fig:E_F_Lz_F}
\end{figure}

In the upper panel of Fig.~\ref{fig:E_F_Lz_F} we show the binary fractions using different cuts in energy, $E^{\mathrm{cut}}$ that separate our sample to high- and low-energy sources (see red and violet points respectively). Our analysis demonstrates that below a critical energy threshold ($E < E^{\mathrm{cut}} \approx E_{\odot}$), binary fractions remain relatively similar. This consistency can be explained by the high contamination rate of \emph{in-situ} stars which dominates the sample of high-energy stars, making the two subsamples of low- and high-energy similar in stellar properties. Conversely, above this energy limit ($E > E^{\mathrm{cut}} \approx E_{\odot}$), we observe a noticeable increase in binary fractions. This trend intensifies at higher energy levels, where the sample likely comprises a purer selection of \emph{accreted} stars. These results align with the findings of our previous section, where we estimated binary fractions in a conservative subsample of \emph{accreted} and \emph{in-situ} stars. The increase in binary fraction at higher energy bins suggests a more significant presence of dynamically heated, \emph{accreted} structures especially of the GS/E, with lower metallicities.


The variation of binary fractions with angular momentum also presents an interesting trend. In the bottom panel of Fig.~\ref{fig:E_F_Lz_F} we show the binary fractions as function of the angular momentum cut-off ($L_z^{\mathrm{cut}}$) that separates our sample to retrograde-prograde and slow-fast rotation sources (see red and violet points respectively). We find consistent higher binary fractions in sources with $L_z < L_z^{\mathrm{cut}}$ compared to the complement subgroup. In stark contrast, the most prograde sources, predominantly associated with the high-$\alpha$ disc and \emph{Splash} \citep{Bonaca17, Gallart2019,diMatteo2019,Belokurov20,Dodd2023}, exhibit significantly lower binary fractions than the overall halo sample. Possible explanations for this difference might include the higher metal content associated with this subsample compared to the slow and retrograde sources. Furthermore, as recently suggested by \citep{Mazzola20}, high abundances of $\alpha$ elements also suppress multiplicity at most values of [Fe/H]. Consequently, the combination of higher metallicity and elevated $\alpha$ element abundances likely contributes to the lower binary fractions observed.

\section{Metallicity dependence}
\label{sec:Metallicityfrac}

In this section, we delve into the relationship between metallicity and binary star fractions while considering samples of \emph{accreted} and \emph{in-situ} halo stars. We used the APOGEE DR 17 catalogue \citep{APOGEE17} to obtain estimates of stellar metallicity and cross-matched our \gaia MS sample with it. To minimise effects by stellar mass on binary fraction and to select a representative sample similar to our halo sample, we limited our selection to sources with APOGEE effective temperature (\texttt{TEFF}) between $4600-6200$ K and stellar mass, approximated using the \cite{Torres10} relation, within the $0.5-1.2M_{\odot}$ range. 

\begin{figure*}
    \centering
    \includegraphics[width=0.48\textwidth]{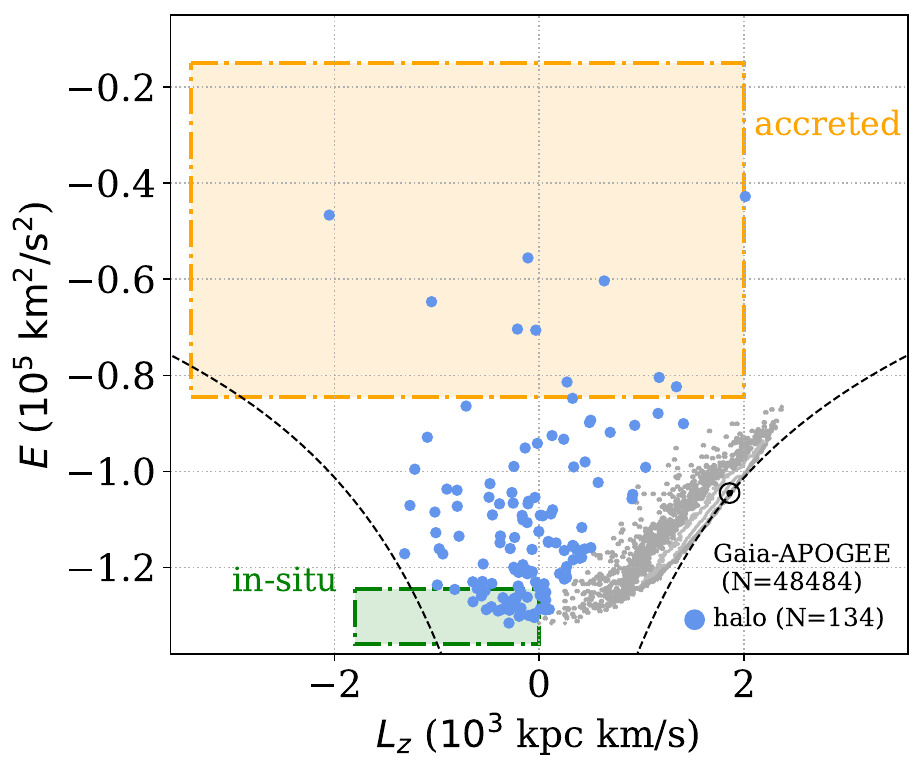}
    \includegraphics[width=0.48\textwidth]{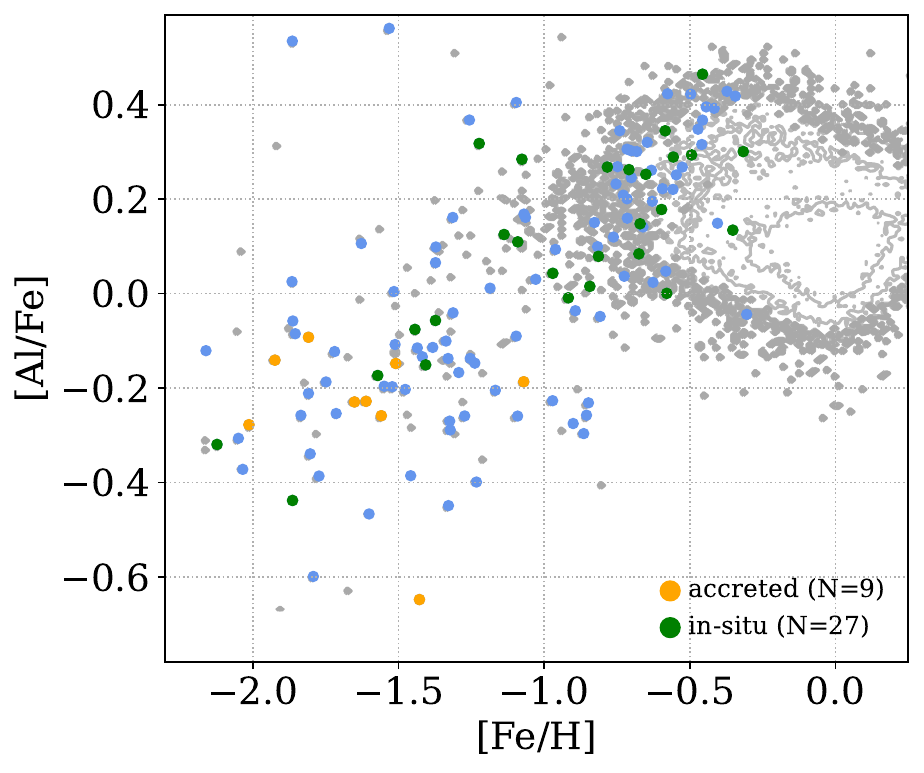}
    \caption{The \textit{Gaia}-APOGEE cross-match sample. Left panel shows the $E-L_z$ values, while the right panel shows the [Fe/H]-[Al/Fe] scatter. Green points, mark \emph{in-situ} stars based on previous section kinematic selection, orange points mark the \emph{accreted} sample. Grey contour background points mark the main location of the disc and other \emph{in-situ} substructures.}
    \label{fig:Gaia-APOGEE}
\end{figure*}

On the one hand, while there are larger catalogues with reported metallicity estimates \citep[e.g.][]{Andrae23}, it is extremely important when dealing with metal-poor binary stars to use a high spectroscopic derivation of the metal content to avoid overestimation of the derived metallicities. On the other hand, performing our binary search analysis solely using APOGEE stars would not be applicable since most targets had only a single exposure with a small fraction of sources having 2-4 RV epoch observations \citep{APOGEE17, Mazzola20}. This limited number of epochs is insufficient to get good estimates of \sigrv~compared to \textit{Gaia}'s DR3 typical $20-30$ epochs \citep{Katz23}. 

As part of our initial filtering, we removed stars with measurements that are believed to be problematic, as indicated by the following general flags: \texttt{VERY\_BRIGHT\_NEIGHBOR}, \texttt{LOW\_SNR}, as well as these element pipeline analysis flags: \texttt{FE\_H\_FLAG}, \texttt{MG\_FE\_FLAG}, \texttt{AL\_FE\_FLAG}. However, we found that even while using these cuts, there were a few cases where the extracted metallicity estimates listed in APOGEE were underestimated. In most of these cases, the outliers strongly correlated with high values of \sigrv, suggesting that the binary nature of these sources, especially in cases of high mass-ratios, have biased the elemental estimation. These cases are common particularly in blended sources and multiple-star systems with more than one star that contribute significantly to the overall flux. A similar issue was also highlighted in \cite{Price-Whelan20}, where, in their analysis aimed at characterising close binaries, it was assumed that their sample consisted exclusively of single-line sources.

Consequently, to remove these outliers, we found the value on \texttt{RV\_CCFWHM}, which marks the Full-Width Half-Maximum (FWHM) of the cross-correlation peak from the combined vs. best-match synthetic spectrum, as a good indicator of `well-behaved' solutions in cases with values below $150$ km/s. The cross-correlation technique is employed to match observed stellar spectra against a library of synthetic spectra. The FWHM of the cross-correlation peak, denoted by \texttt{RV\_CCFWHM}, quantifies the width of this peak at half of its maximum height, reflecting the degree of match between the observed and synthetic spectra. Thus, a peak that is too broad might indicate an imprecise match. 

The cross-match of the \textit{Gaia}-APOGEE sample, yielded $N=48,484$ sources (disc and halo) in common, with only $N=9$ and $N=27$ \emph{accreted} and \emph{in-situ} sources respectively having using section~\ref{sec:haloSample} subsamples selection.
In the left panel of Fig.~\ref{fig:Gaia-APOGEE}, we show a scatter plot of the sources on the $E-L_z$ plane similar to Fig.~\ref{fig:Lz_E} while on the right panel we show the distribution on the [Fe/H]-[Al/Fe] diagram. As expected, our kinematic selected halo sub-samples dominate the metal-poor fraction of stars, with \emph{accreted} sources having lower [Al/Fe] values compared to \emph{in-situ} stars \citep{Hasselquist21, BelokurovKravtsov22,BelokurovKravtsov23, DeasonBelokurov24}. In addition, we note on a few other \emph{in-situ} metal-poor prograde stars seen in this plot (grey points), probably affiliated with the \emph{Aurora} population, which while assumed to be kinematically hot, has an approximately isotropic
velocity ellipsoid and a modest net rotation \citep{BelokurovKravtsov22, Zhang24}.

Using the \textit{Gaia-}APOGEE sample, we can attempt to estimate binary fractions in bins of metallicity. The black crosses in Fig.~\ref{fig:feh_F} present a detailed analysis of the binary fraction in various metallicity bins. We focus on the median values and the $16\%$, $84\%$ quantile values of the binary fractions posteriors. In addition we plot by red and violet crosses the fraction of retrograde ($L_z <0$; $N=51$) and prograde ($L_z \geq 0$; $N=83$) halo sources respectively.

The low number of \emph{accreted} ($N=9$) and \emph{in-situ} ($N=27$) halo stars in our \textit{Gaia}-APOGEE cross-matched subsample is not sufficient to produce valuable estimates of binary fractions using our Bayesian model. Instead, we decided to estimate the median and $16\%$, $84\%$ quantile values on the metallicty values in each halo subsample and place the binary fractions found in Section~\ref{sec:halofrac} as the expected values. Consequently, we mark by orange and green boxes the \emph{accreted} and \emph{in-situ} binary fractions to stress that these are not a direct calculation of the binary fractions of corresponding samples shown in Fig.~\ref{fig:Gaia-APOGEE}.  

Examining the trend in field stars' binary fraction as a function of [Fe/H], Fig~\ref{fig:feh_F} corroborates previous research findings that have consistently shown higher binary fractions in metal-poor stars \citep{Badenes18, Moe19, Mazzola20, Price-Whelan20}. This trend is prominently visible across the different metallicity bins, reinforcing the significant role of metallicity in the formation of binary stars. 

In our context, one further notable observation from Fig.~\ref{fig:feh_F} is the
consistency of the \emph{accreted} and \emph{in-situ} with the field stars trend, suggesting that the anti-correlation in close-binary fractions with metallicity continues into the domain of halo stars. Secondly, 
we find an interesting discrepancy in field binary fractions compared to the halo prograde stars. A similar discrepancy is also evident when comparing the trend of field binary fractions in Fig.~\ref{fig:feh_F} to the highest prograde bin ($L_z \geq L_z^{\mathrm{cut}}=0.5$) in the bottom panel of Fig.~\ref{fig:E_F_Lz_F}.
Assuming that the prograde bin includes high-$\alpha$ disc and \emph{Splash} stars, we might expect it to be populated mainly by sources with [$\alpha$/Fe] $\geq 0.2$.
However, our bin estimate of a binary fraction of $F\approx0.10$, is lower than the expected value observed in the trend of Fig.~\ref{fig:feh_F} for similar metallicity. Nonetheless, given the \textit{Gaia}-APOGEE sample is dominated by thin disc stars, most of the sources in this diagram have lower $\alpha$ values and have a higher stellar mass compared to thick disc stars. Both of these parameters inspire a higher binary fraction \citep{Raghavan10, Moe19, Mazzola20} that can mitigate this effect. 

\section{Conclusions}
\label{sec:Conclusions}
This study provides valuable insight into close binary star fractions within the Galactic \emph{accreted} and \emph{in-situ} halo populations, which stands as a pivotal component in understanding the dynamical history and formation of the Milky Way Galaxy. 

By leveraging the comprehensive data from Gaia DR3 \citep{Vallenari23}, our methodological approach, utilised a novel Bayesian framework to analyse the radial velocity (RV) uncertainty distribution of \gaia RVS sources \citep{Katz23, Sartoretti23}. Using this approach, we were able to model the populations of single and binary stars as a sum of two Gaussian distributions, allowing a nuanced understanding of close-binary star fractions in different Galactic environments using only \gaia data.

Our findings reveal a higher binary fraction in \emph{accreted} stars, compared to \emph{in-situ} halo stars. On the one hand, this observation suggests that the processes that influence binary star formation and evolution might differ across these distinct populations. On the other hand, given the slightly lower metal content in \emph{accreted} stars compared to \emph{in-situ} halo stars, the difference in binary fractions might be explained by the difference in metallicity. 
A similar conclusion was also reached for wide-binary fractions in \cite{Hwang22}, although their selection of \emph{accreted} and \emph{in-situ} sources was different from the one we used. Furthermore, their selection of stars on the $E-L_z$ plane was more contaminated, causing the binary fractions to be much more consistent in any case. 

\begin{figure*}
	\includegraphics[width=13.5cm]{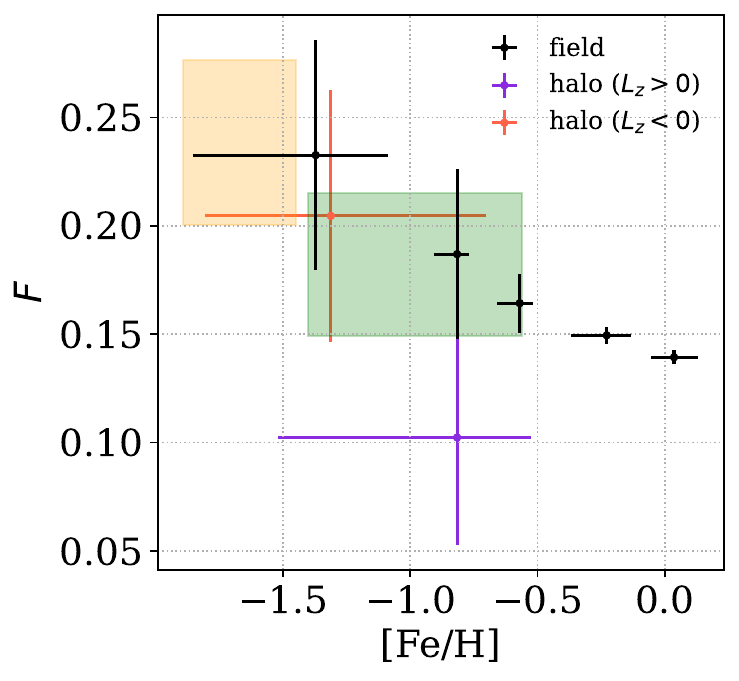}
    \caption{Posterior values of binary fractions using APOGEE data in bins of [Fe/H]. Black crosses mark the \textit{Gaia}-APOGEE sample. Red and violet crosses mark the fractions in retrograde ($L_z <0$) and prograde ($L_z \geq 0$) halo sources respectively. Orange and green boxes mark the \emph{accreted} and \emph{in-situ} binary fractions using on the values estimated in Section~\ref{sec:halofrac}.}
    \label{fig:feh_F}
\end{figure*}

In our study, we also explored binary fractions across different galactic components, employing varied energy and angular momentum cuts to differentiate these groups. We found that, below a specific energy threshold ($E < E^{\mathrm{cut}} \approx E_{\odot}$), binary fractions remained relatively uniform due to a high contamination rate of \emph{in-situ} stars that dominate the high-energy sample \citep{Naidu20,BelokurovKravtsov23,Zhang24,DeasonBelokurov24}, resulting in similar stellar properties between low- and high-energy stars. Above this energy threshold, binary fractions increased, especially at higher energy levels where the sample likely consists more of \emph{accreted} stars, of the GS/E, indicating a prevalence of dynamically heated metal-poor structures. Additionally, when examining the variation of binary fractions with angular momentum, we observed consistently higher binary fractions in retrograde sources ($L_z < L_z^{\mathrm{cut}}$) compared to their prograde counterparts. This disparity is stark, as the most prograde sources, which are typically metal-rich, exhibited significantly lower binary fractions than those in the overall halo sample. These findings suggest that the observed trends in binary fractions can provide insights into the dynamic and compositional differences within the Galaxy.

In general, our research corroborates previous studies, indicating higher binary fractions in metal-poor stars \citep{Moe19, Price-Whelan20, Mazzola20}. Our analysis reveals that the binary fractions of \emph{accreted} and \emph{in-situ} samples are consistent with the field stars trend, as evidenced in Fig.~\ref{fig:feh_F}. This observation suggests that the anti-correlation between close-binary fractions and metallicity continues into the domain of halo stars. Particularly, it underscores the influence of metallicity during the early stages of the Milky Way's formation and in the \emph{accreted} dwarf galaxies. The higher prevalence of binary systems in metal-poor environments suggests a fundamental aspect of binary formation in the early universe, where lower metallicity was more common. Such subtleties in the data not only enhance our understanding of the binary star dynamics but also pose intriguing questions about the historical processes that have shaped these populations.



Building on this, the impact of primary mass on binary fraction adds another layer of complexity to our analysis. Since the binary fraction depends on the primary mass of the system \citep{Raghavan10, Moe19}, exploring this relationship could bias our results. To address this, we conducted a comparative analysis of the Absolute Magnitude (Abs. G) distributions between the two halo samples, hypothesizing that it could serve as a proxy for primary mass. Our findings revealed that the distributions in Abs. G between the \emph{accreted} and \emph{in-situ} samples were comparable, suggesting that despite the challenges posed by metallicity, the primary mass, as inferred from Abs. G., does not significantly differentiate the binary fractions of the two halo samples. Similarly, in the analysis of the \textit{Gaia}-APOGEE metallicity bins we intentionally restricted our sample to narrow region of effective temperature
$4600-6200$ K and stellar mass within the $0.5-1.2M_{\odot}$ range to decrease possible biases with primary mass. However, this in any case would not change dramatically, nor reverse, the observed trend of higher binary fraction towards metal-poor stars. In fact, as \cite{Moe19} pointed out, as metal-poor stars tend to be of later type, the trend seen of higher binary fraction in metal-poor stars should become more significant.

The latest discoveries of dormant compact objects, such as Gaia BH3 \citep{BH3} and a list of Neutron star candidates \citep{El-BadryNeutron24}, have highlighted the critical importance of exploring binary fractions in metal-poor regions, particularly among halo stars. These findings suggest that metal-poor environments, often associated with older galactic structures, may harbor a significant number of these elusive compact remnants. The presence of such objects in binary systems can provide valuable insights into the evolutionary paths that lead to their formation.

Future work could aim to expand the halo sample utilised in this study to obtain more accurate estimates of the mean values of binary fractions and their uncertainties. For instance, we limited our analysis to stars with \grvs<12 to ensure the reliability and quality of the RV data. Nonetheless, the \gaia RVs are available to stars up-to \grvs=14. Another potential strategy involves leveraging the RV scatter, incorporating data from various ground-based spectroscopic surveys such as LAMOST \citep{Cui12}, GALAH \citep{GALAH15} and APOGEE \citep{MajewskiAPOGEE17}. This approach would allow us to extend our sample, including a broader range of stars suspected to be in binary systems.

In terms of extending our methodical approach, future works might also exploit our Bayesian model to estimate the likelihood of individual \gaia sources being affiliated as single stars or as binary stars. 
This refined focus could significantly enhance our understanding of the distinct characteristics and behaviours of specific systems allowing further study of their specific nature.

\section*{Acknowledgements}
We would like to thank the reviewer for their helpful comments and suggestions. D.B. acknowledges the support of the Blavatnik family, the British council, and the British Friends of the Hebrew University (BFHU) as part of the Blavatnik Cambridge Fellowship. D.B. also extends his gratitude to Didier Queloz for his hospitality and encouragement. We wish to thank Shay Zucker for a helpful discussion on the Bayesian methodology and Jo Bovy for assisting with the definition of the \texttt{Galpy}'s galactic potential. This work has also made use of data from the European Space Agency
(ESA) mission \gaia (https://www.cosmos.esa.int/gaia), processed
by the \gaia Data Processing and Analysis Consortium (DPAC; ht
tps://www.cosmos.esa.int/web/gaia/ dpac/consortium). Funding for
DPAC has been provided by national institutions, in particular the
institutions participating in the \gaia Multilateral Agreement. 
Funding for the Sloan Digital Sky Survey IV has been provided by the Alfred P. Sloan Foundation, the U.S. Department of Energy Office of Science, and the Participating Institutions. SDSS acknowledges support and resources from the Center for High-Performance Computing at the University of Utah. The SDSS web site is www.sdss4.org.

SDSS is managed by the Astrophysical Research Consortium for the Participating Institutions of the SDSS Collaboration including the Brazilian Participation Group, the Carnegie Institution for Science, Carnegie Mellon University, Center for Astrophysics | Harvard \& Smithsonian (CfA), the Chilean Participation Group, the French Participation Group, Instituto de Astrofísica de Canarias, The Johns Hopkins University, Kavli Institute for the Physics and Mathematics of the Universe (IPMU) / University of Tokyo, the Korean Participation Group, Lawrence Berkeley National Laboratory, Leibniz Institut für Astrophysik Potsdam (AIP), Max-Planck-Institut für Astronomie (MPIA Heidelberg), Max-Planck-Institut für Astrophysik (MPA Garching), Max-Planck-Institut für Extraterrestrische Physik (MPE), National Astronomical Observatories of China, New Mexico State University, New York University, University of Notre Dame, Observatório Nacional / MCTI, The Ohio State University, Pennsylvania State University, Shanghai Astronomical Observatory, United Kingdom Participation Group, Universidad Nacional Autónoma de México, University of Arizona, University of Colorado Boulder, University of Oxford, University of Portsmouth, University of Utah, University of Virginia, University of Washington, University of Wisconsin, Vanderbilt University, and Yale University.
This research also made use of TOPCAT \citep{Taylor05}, an interactive graphical viewer and editor for tabular data. 

This research has made use of the NASA Exoplanet Archive, which is operated by the California Institute of Technology, under contract with the National Aeronautics and Space Administration under the Exoplanet Exploration Program.

This research has made use of the SIMBAD database, CDS, Strasbourg Astronomical Observatory, France.

\section*{Data Availability}

The research presented in this article predominantly relies on data that is publicly available and accessible online through the \gaia~DR3 and APOGEE archives.



\bibliographystyle{mnras}
\bibliography{main} 





\bsp	
\label{lastpage}
\end{document}